\documentclass{epl}
\usepackage{graphicx}

\title{Observation of phonon structure in electron density of states
of a normal metal}
\author{A. Knigavko\inst{1} \and J.~P. Carbotte\inst{1} \and F. Marsiglio\inst{2}}
\institute{
  \inst{1} Department of Physics and Astronomy, McMaster University, Hamilton,
  Ontario, Canada, L8S 4M1\\
  \inst{2} Department of Physics, University of Alberta, Edmonton, Alberta,
  Canada, T6J 2J1
}

\begin{document}

\maketitle

\begin{abstract}
Within standard treatments of the interacting electron--phonon
system the electron density of states (EDOS) shows no sign of the
phonons when the system is in the normal state. On the other hand,
emergence of a fine energy scale in the superconducting state makes
the phonons readily observable in the EDOS and other spectroscopic
quantities. Here we wish to re-examine the conditions under which
phonon structure might be present in the normal state. Accounting
for a finite, i.e. non-infinite, electronic bandwidth results in
easily distinguishable phonon structure in the EDOS. We argue that
for narrow band metals, the resolution of photoemission spectroscopy
is sufficient for resolving these structures. Manifestations of
finite band effects in $K_{3}C_{60}$ are discussed.
\end{abstract}

%%%%%%%%%%%%%%%%%%%%%%%%%%%%%%%%%%%%%%%%
\section{Introduction}
The observation of phonon structure in tunneling data for
superconducting materials became a very important diagnostic tool in
the 1960's \cite{mcmillan69}. Following this initial success
attempts were made (with some degree of success) to extract similar
information in the normal state \cite{steinrisser68,chen70,wolf85}.
Part of the difficulty at the time no doubt was due to the fact that
calculations using many-body Green functions predicted that no
phonon structure should be present in the Electronic Density Of
States (EDOS) \cite{englesberg63,mars-book}. When the bare EDOS is
infinitely wide and featureless and when particle-hole symmetry
is present, the dressed EDOS retains its constant bare band value
and the phonon signatures drop out.  On the other hand, as was
studied extensively in the 1980's (see for instance
\cite{mitrovic83,pickett82}), phonon renormalizations do enter when
the EDOS has significant energy dependence near the Fermi surface on the
scale of phonon energies.

More recently
\cite{alexandrov87,yoo00,cappelluti01,cappelluti03,dogan03,knigavko05}
it was realized that phonon structure can also arise when a finite
band cutoff is applied in the EDOS on the scale of the bandwidth. In
this letter we want to accomplish two goals: first we want to
understand under what conditions these structures can be resolved in
Angle Integrated Photo Emission Spectroscopy (AIPES) and/or
tunneling even for relatively broad bands, and secondly, we wish to
establish the scale of the structures for parameters more specific
to a real narrow band material, in this case K$_3$C$_{60}$. While
our primary interest here is in the EDOS we also provide comparison with
other spectroscopic quantities such as the real and imaginary parts
of the quasiparticle self energy and its optical counterpart, the
memory function, both of which have phonon signatures even when the band
is assumed to be infinite. It is well known that even when these
structures are not directly visible, as in the optical scattering
rate, they can be emphasized either by differentiation
\cite{marsiglio98}, or extracted by other more sophisticated
``inversion'' procedures \cite{marsiglio98,dordevic05}. In contrast,
the phonon
structures due to a finite bandwidth are more readily detectable.

To see sizeable features in the EDOS, the electron-phonon
interaction must be sufficiently strong. To be specific we consider
the Eliashberg $\alpha ^{2}F(\omega)$ function appropriate to
the intermediate coupling strength superconductor
K$_3$C$_{60}$. We start with a large bare bandwidth $W = 2.5$~eV
and find that for the clean system and at low temperature the phonon
structure in the renormalized EDOS is very prominent in the range of
phonon energies (keep in mind that for an infinite bandwidth the
structure is absent).
Furthermore, the presence of impurity scattering does not obliterate
the structure; it merely reduces its size \cite{cappelluti03}.
Reducing the bare bandwidth to $W = 0.5$~eV, a value more
relevant for K$_3$C$_{60}$, we find that all dominant features of
the renormalized EDOS are determined by the electron--phonon
interaction.

%%%%%%%%%%%%%%%%%%%
\section{Bare EDOS, the self energy and the renormalized EDOS}
We use the following model for the bare electronic band:
$
N_{0}(\xi )=N_{0}\Theta \left( W/2-|\xi |\right) ,
$
where $W$ is the bare band width and $\Theta (x)$ is the step function.
The constant $N_{0}$ is fixed by normalization: $N_{0} = 1/W$.
In this letter we retain particle-hole symmetry for simplicity, with the
chemical potential at the center of the band, $\mu=0$.

The electronic self energy $\Sigma (z)=\Sigma _{1}(z)+{\rm i}\Sigma _{2}(z)$
is calculated from the Migdal equations formulated in the
mixed real--imaginary axis representation \cite{cappelluti03,marsiglio88}:
\begin{eqnarray}
\Sigma (z) &=&\Gamma \, \eta (z)+T\sum_{m=-\infty }^{+\infty }
\lambda(z-{\rm i}\omega _{m})\eta ({\rm i}\omega _{m})
\nonumber \\
&+& \int_{0}^{\infty }{\rm d}\omega \,\alpha ^{2}F(\omega )\left\{ \left[
f(\omega -z)+n(\omega )\right] \eta (z-\omega )
+ \left[ f(\omega +z)+n(\omega )\right] \eta (z+\omega )\right\} ,
\label{self-energy} \\
\lambda (z) &=&\int_{0}^{\infty }{\rm d}\omega \,\alpha ^{2}F(\omega )
\frac{2\omega }{\omega^{2}-z^{2}} ,
\label{lambda} \\
\eta (z) &=&\int_{-\infty }^{\infty }{\rm d}\xi \,\frac{N_{0}(\xi )}{N_{0}(0)}%
\frac{1}{z-\xi -\Sigma (z)} ,
\label{eta}
\end{eqnarray}
where $\omega _{m}=\pi T(2m-1),$ $m\in Z$ are the fermionic Matsubara
frequencies, and $f(\omega )$ and $n(\omega )$ are the Fermi and Bose distribution
functions respectively. The electron--phonon interaction is specified in
terms of the electron--phonon spectral function $\alpha ^{2}F(\omega )$ (the
Eliashberg function). The parameter $\Gamma $, which has the meaning of an impurity
scattering rate, specifies the strength of the interaction with impurities.
Note that finding $\Sigma (z)$ requires a self consistent solution of
eqs.~(\ref{self-energy})--(\ref{eta}), which is a consequence of accounting for
the finite bare band width in eq.~(\ref{eta}).

The variable $z$ can assume arbitrary complex values. At first, the
solutions for $\Sigma (z)$ are sought on the imaginary axis, at
$z={\rm i}\omega _{m}$, where eq.~(\ref{self-energy}) is simpler.
Then, the function $\eta ({\rm i}\omega _{m})$ is used to set up an iterative
procedure to find $\Sigma (\omega )$ just above the real axis
(we use $\omega $ as shorthand for $\omega+{\rm i}0^{+}$).
While we will focus on the EDOS, which can be determined in AIPES experiments,
the real axis self energy, both its real and imaginary parts, can be
obtained from the angular-resolved photoemission. The accuracy of this technique
has increased dramatically in recent years and properties of both new and
traditional materials have been scrutinized (see, for example,
refs.~\cite{valla99,chainani00,reinert03,devereaux04}
for a flavour of the recent developments).

Motivated by the electron--phonon interaction in the fulleride compound
K$_{3}$C$_{60}$, we use a three frequency model for the electron-phonon
spectral function:
\begin{equation}
\alpha ^{2}F(\omega )=\lambda \sum_{i=1}^{3}\frac{\omega _{i}l_{i}}{2}\delta
\left( \omega -\omega _{i}\right),
\quad\quad
\sum_{i=1}^{3}l_{i}=1
\label{three-freq-model}
\end{equation}
with $l_{1}=0.3,l_{2}=0.2,l_{3}=0.5$ and
$\omega _{1}:\omega _{2}:\omega_{3}=0.04:0.09:0.19$ eV \cite{choi98}.
The interaction strength $a$ is defined as the area under the
$\alpha ^{2}F(\omega )$ curve. The mass enhancement parameter $\lambda $
is given by eq.~(\ref{lambda}) with $z=0$.
For the forthcoming discussion we set $\lambda =0.71$. Then this model has
$a=43.8$ meV and $\omega _{\ln }=102.5$ meV, where $\omega _{\ln }$ is the
logarithmic frequency \cite{mars-book}, a convenient parameter to quantify
the phonon energy scale. In the last section we consider different values
of $\lambda$ as well. Finally, below we use $W=2.5$ eV.

The renormalized density of electronic states (or density of states for
quasiparticles) is defined by
\begin{equation}
N(\omega )=\int_{-\infty }^{+\infty }{\rm d}\xi N_{0}(\xi )A(\xi,\omega ),
\label{dos-renorm}
\end{equation}
where $A(\xi ,\omega )=-\mathop{\rm Im}G_{ret}(\xi ,\omega )/\pi $ is the
electronic spectral density and $G_{ret}(\xi ,\omega )$ is the retarded
Green function. It can be expressed in terms of the function $\eta =\eta
_{1}+{\rm i}\eta _{2}$ \ of Eq. (\ref{eta}) through $N(\omega
)/N_{0}(0)=-\eta _{2}(\omega )/\pi$. In AIPES the measured intensity,
$I(\omega )$, is the product of the renormalized EDOS
and a thermal factor $f(\omega )$
which provides a cutoff at $T=0$, convoluted with the instrument
resolution $R(\omega )$ which can be taken as a Gaussian with half-width
of $2.8$ meV~\cite{reinert03}. With particle--hole symmetry
it is convenient to define a symmetrized quantity
$I_{s}(\omega )=\left[ I(\omega )+I(-\omega )\right] /2$ from which
$f(\omega)$ drops out \cite{norman-nature}, leading to
\begin{equation}
I_{s}(\omega )=\int_{-\infty }^{\infty }{\rm d}\omega'
N(\omega ^{\prime })R(\omega ^{\prime}-\omega ).
\label{I-sym}
\end{equation}
%%%%%%%%%%%%%%%%%%%%%%%%%%%%%%%%%%%%%%%
\begin{figure}
\twofigures[width=6.5cm]{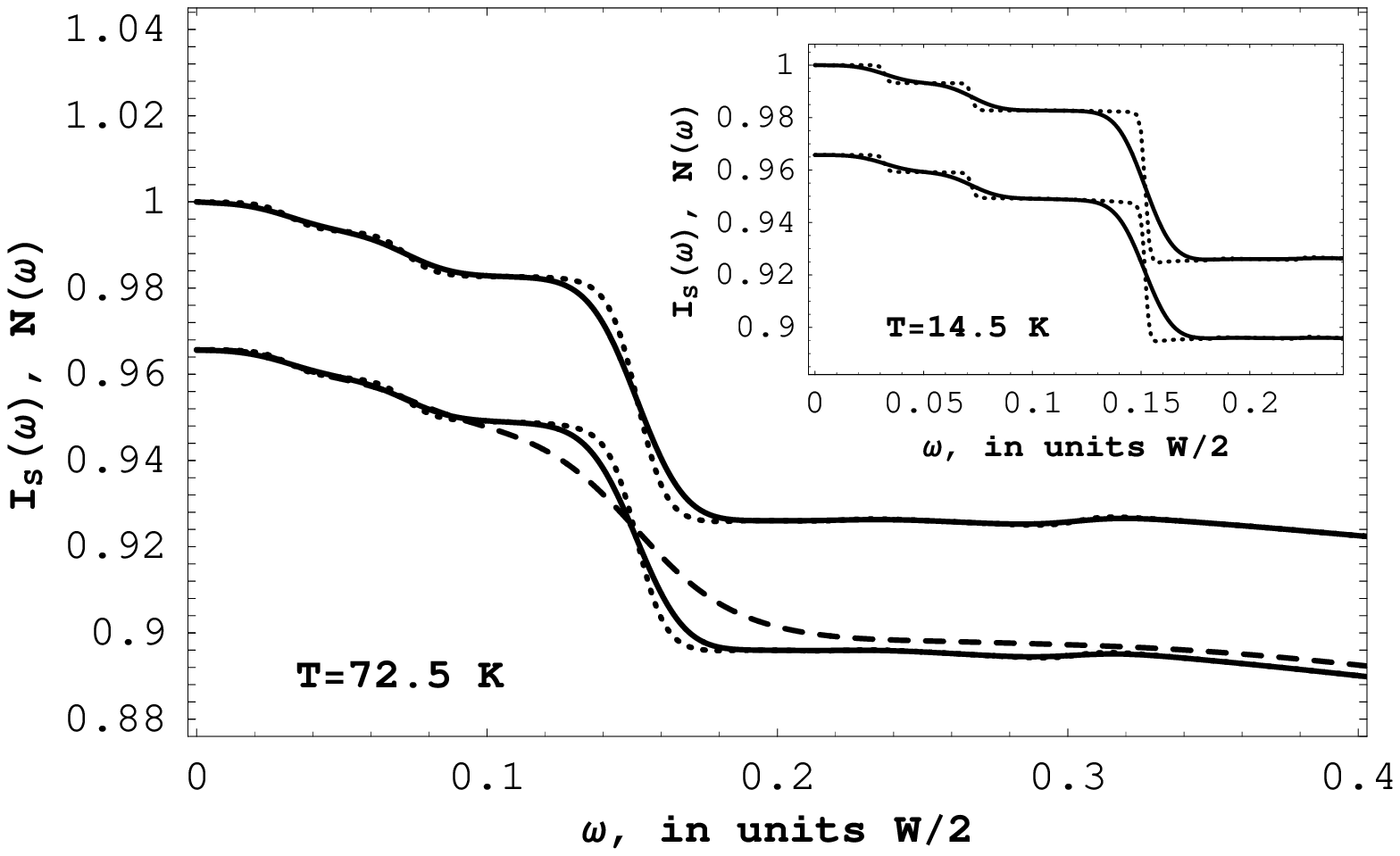}{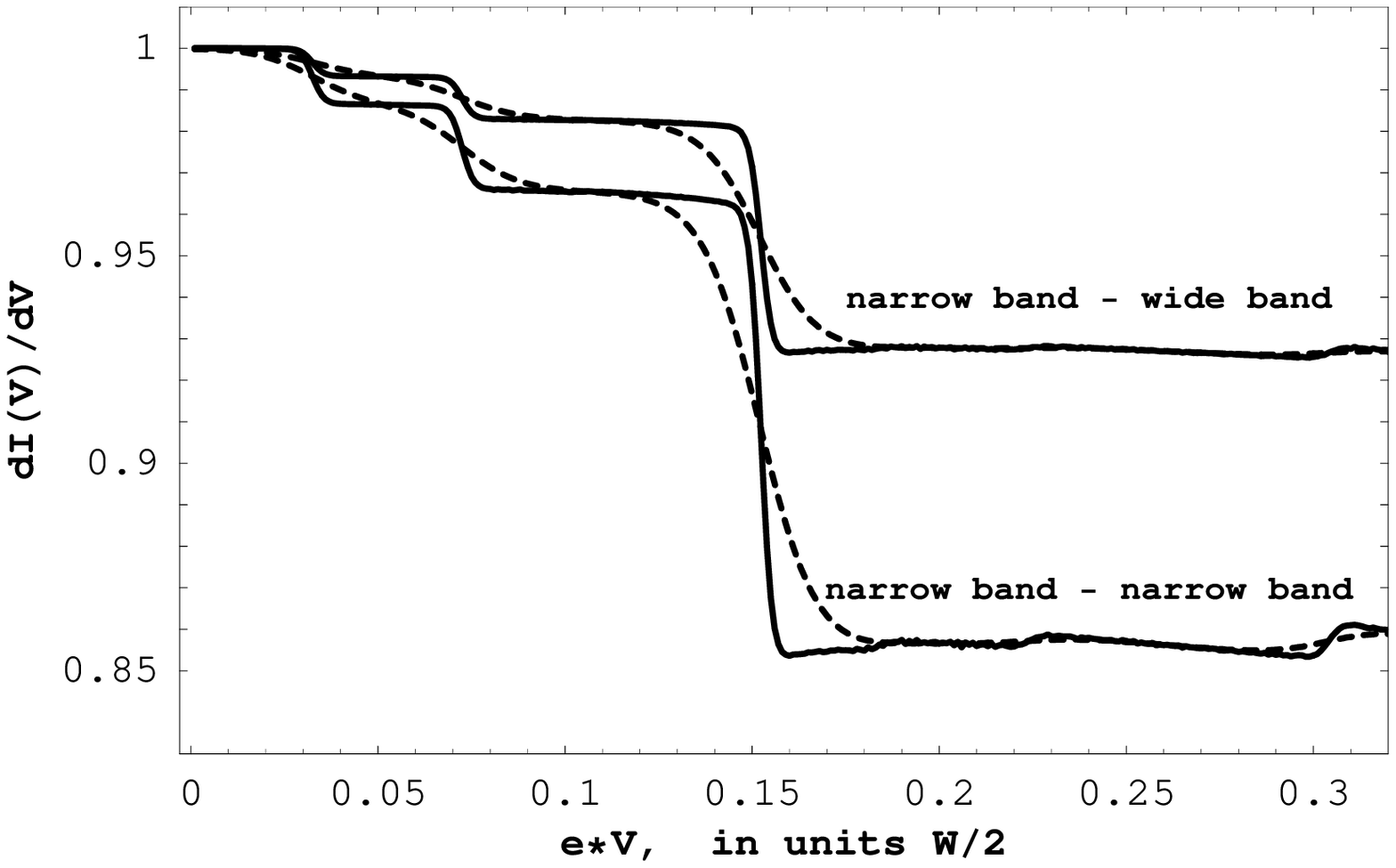}
\caption{
Comparison of the renormalized density of states $N(\omega )$ (dotted) with
the symmetrized photoemission signal $I_{s}(\omega)$ (solid) that includes the
instrument resolution for $\alpha^2F(\omega)$ of eq.~(\ref{three-freq-model}).
The impurity scattering rate is $\Gamma=0$ and $22$ meV for the upper and lower
pairs of curves respectively. Frequency $\omega$ is given in units of half bare
bandwidth $W/2=1.25$ eV. The temperature is $T=72.5$ K.
For a representative extended $\alpha^2F(\omega)$ the results (dashed curve) for
$N(\omega )$ and $I_{s}(\omega)$ are indistinguishable on the scale of the plot.
Inset: the same as the main frame but for $T=14.5$ K.
}
\label{fig-dos-resolut}
\caption{
Differential I--V chacteristic for tunelling between
two identical narrow band (i.e. W = 2.5 eV) metals (lower two curves) and between
narrow and wide band metals (upper two curves).
The temperature is $T=14.5$ K (solid curves) and $72.5$ K (dashed
curves), and the impurity scattering rate is $\Gamma=22$ meV.
%Evergy difference between contacts $e*V$ is given
%in units of half bare bandwidth $W/2=1.25$ eV.
}
\label{fig-dos-iv}
\end{figure}

In fig.~\ref{fig-dos-resolut} we show results (solid curves) for
$I_{s}$ vs $\omega $ and compare with the density of states $N(\omega )$
(dotted curves). The main frame is for $T\simeq 72.5K$ while the inset
is for $T=14.5$ K. Two different scattering rates are used, namely
the clean case $\Gamma =0$ (top curves) and $\Gamma =a/2\approx22$ meV
(bottom curves). For the clean case the curves start at value unity
at zero frequency. The introduction of impurity scattering reduces the value
of the density of states in the low frequency range shown in
fig.~\ref{fig-dos-resolut} but
does not significantly smear the predicted phonon structure. Three drops are
clearly seen at the frequencies of each of the three delta functions in the
model spectral density $\alpha ^{2}F(\omega )$ of
eq.~(\ref{three-freq-model}).
%Had we used an extended spectrum instead the three steps in $N(\omega)$
%would be smoothed out but the general downward trend would continue.
We note that increasing the impurity scattering does
not smear out the predicted phonon structure, rather it reduces the
effective step drop. On the other hand, comparison of the inset with the
main frame shows that temperature does smooth out the curves considerably,
but even for the worst case $T = 72.5$ K and $\Gamma = 22$ meV, the structure in
$I_{s}(\omega)$ remains. For the chosen parameters, which are not extreme,
instrument resolution does not have a strong smearing effect.
We also used a representative extended $\alpha ^{2}F(\omega )$ with the same
$\lambda$ and position of the peaks as in eq.~(\ref{three-freq-model}). In this case
the results (dashed curve) for $N(\omega )$ and $I_{s}(\omega)$ are
indistinguishable on the scale of fig.~\ref{fig-dos-resolut}.

%%%%%%%%%%%%%%%%%%%%%%%%%%%%%%%%%%%%%%%%%%%
%\section{Tunneling}
Another way to detect the finite band phonon structures in the EDOS is through
tunneling spectroscopy. The well known formula for the current, $I$ (assuming
the tunneling matrix element is independent of energy), through a tunnel
junction reads:
\begin{equation}
I(V)\sim \int_{-\infty }^{+\infty }{\rm d}\xi \frac{N_{L}(\xi )}{N_{L}(0)}\frac{%
N_{R}(\xi )}{N_{R}(0)}\left[ f\left( \xi \right) -f(\xi +eV)\right] ,
\label{tun-current}
\end{equation}
where $V$ is the applied voltage, $e$ is the charge of the electron and the
labels $L$ and $R$ denote materials on opposite sides of the junction.
The normalization of $N_{L,R}(\omega )$ is to its value at $\omega =0$.
In fig.~\ref{fig-dos-iv} we show  ${\rm d}I(V)/{\rm d}V$ vs $e*V$ for
a narrow band  -- wide band case (upper pair of curves) and for the case where
the same narrow band metal is on both sides (lower pair of curves). The impurity
scattering rate is $\Gamma=a/2\approx22$ meV and the two temperatures shown are
$T=14.5$ K and $72.5$ K. In both cases the phonon structures are clearly
identifiable in the differential conductance even at higher temperatures.
The upper curves are almost identical to normalized EDOS $N(\omega )/N(0)$
(see inset in fig.~\ref{fig-dos-resolut}). Features in the lower curves are
about twice as large in magnitude as in the upper curves. Finite band effects
are enhanced when the same narrow band metal is used on each side of tunnel
junction.

%%%%%%%%%%%%%%%%%%%%%%%%%
\section{Optical response}
For characterization of the optical response the quantity of interest is
the memory function.
Its real and imaginary parts are $\tau _{op}^{-1}(\omega )$ and
$-\omega \lambda _{op}(\omega )$ respectively, where $\tau _{op}^{-1}$
is the optical scattering rate and $\lambda _{op}$ is the optical mass
renormalization \cite{mars-book}. The memory function is connected
to the optical conductivity $\sigma =\sigma _{1}+ {\rm i}\sigma _{2}$
by algebraic relations:
$\tau _{op}^{-1}=(2S/\pi)\Re[1/\sigma]$ and
$-\omega \lambda _{op} = \omega -(2S/\pi)\Im[1/\sigma]$,
where $S$ is the total optical spectral weight defined as
$S=\int_{0}^{+\infty }{\rm d}\omega\sigma _{1}(\omega )$.
The optical conductivity for our model was obtained using linear
response theory, neglecting vertex corrections. The result for the real
part reads:
\begin{eqnarray}
\sigma _{1}(\omega )
&=&
\frac{2\pi e^{2}}{\hbar ^{2}}
\int_{-\infty }^{+\infty}{\rm d}\xi N_{0}(\xi )v_{\xi }^{2}
\int_{-\infty }^{\infty }{\rm d}\omega^{\prime }
A\left( \xi,\omega^{\prime }\right)A\left( \xi ,\omega^{\prime } +\omega \right)
\frac{f\left( \omega^{\prime } \right) -f(\omega^{\prime } +\omega )}{\omega },
\label{cond-real-part}
\end{eqnarray}
where $v_{\xi }^{2}$ is the averaged square of the group velocity (see
Ref.~\cite{marsiglio90} for details). We assume that the system is isotropic
and use for $v_{\xi }^{2}$ the expression
$v_{\xi }^{2}=\frac{2\hbar ^{2}}{mD}\left( \frac{W}{2}+\xi \right)$,
derived from the quadratic dispersion of
free electrons with lower band edge at $\xi =-W/2$.
Here $D$ is the number of spatial dimensions, $m$ the free electron mass.
Since the complex conductivity $\sigma(\omega )$ satisfies the
Kramers--Kronig relation, the required imaginary part can be obtained as
the Hilbert transform of the real part.
%%%%%%%%%%%%%%%%%%%%%%%%%%%%%%%%%%%%%%%%%
\begin{figure}
\twofigures[width=6.5cm]{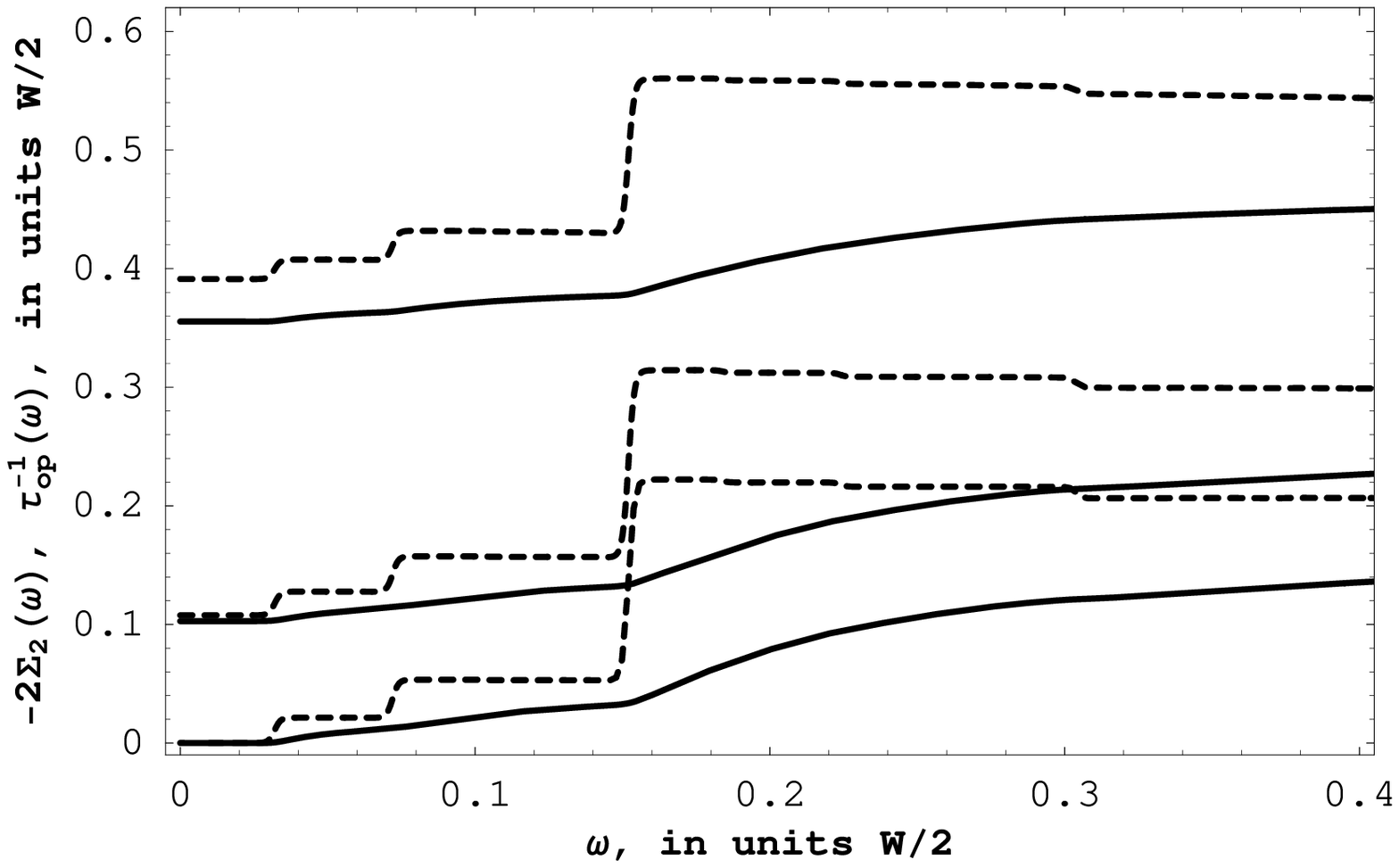}{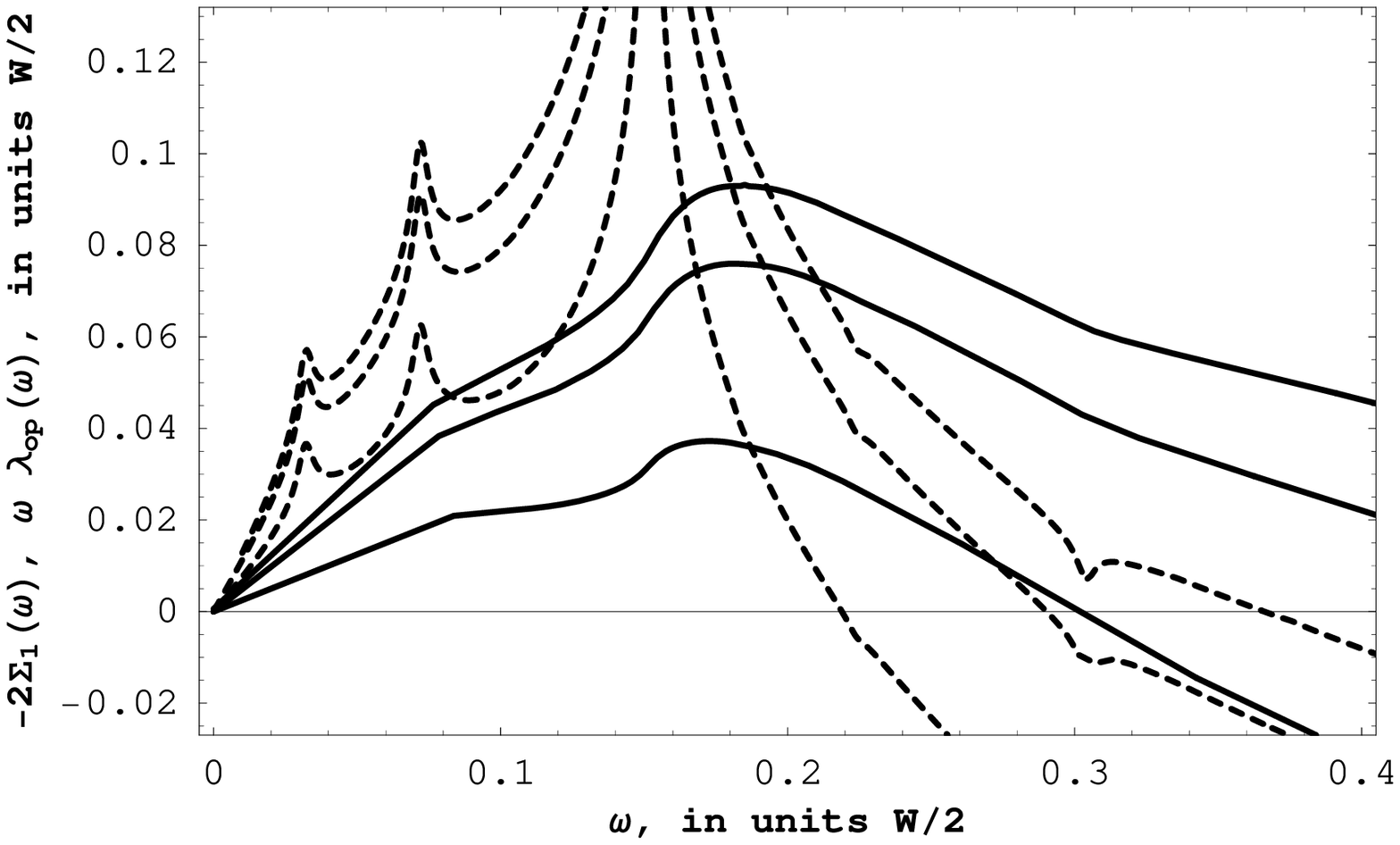}
\caption{
Comparison of the quasiparticle scattering rate $-2\Sigma
_{2}(\omega )$ (dashed) with its optical counterpart $\tau _{op}^{-1}$ $%
(\omega )$ (solid) for impurity scattering rates
$\Gamma =0$, $22$, and $88$ meV, from bottom to top.
The temperature is $T=14.5$ K, the bare bandwidth is $W=2.5$ eV.
}
\label{fig-tau-self2}
\caption{
Comparison of $-2\Sigma _{1}(\omega )$  (dashed) with
$\omega \lambda _{op}(\omega )$ (solid) for the same
values of $\Gamma$ as in fig.~\ref{fig-tau-self2}, but now from top to bottom.
The temperature is $T=14.5$ K, the bare bandwidth is $W=2.5$ eV.
}
\label{fig-omlam-self1}
\end{figure}

The memory function is the optical counterpart of the self energy.
In fig.~\ref{fig-tau-self2} we compare $\tau_{op}^{-1}(\omega )$
with $-2\Sigma _{2}(\omega )$ in the limited range
of frequencies $\omega \in [0,500]$ meV for $T = 14.5$ K.
In the infinite band case these quantities are equal at $\omega=0$.
The bottom pair of curves in fig.~\ref{fig-tau-self2} is for the clean case,
the middle pair is for $\Gamma =a/2\approx 22$ meV,  while the top pair is for
$\Gamma =2a\approx88$ meV. The effect of increasing the impurity content is to
push the curves up corresponding to a larger value of the residual scattering.
Beyond this effect there are changes in phonon structures, but little smearing.
The structures are much more visible in the quasiparticle scattering rate
$-2\Sigma _{2}$ (steps) than in the corresponding optical quantity $\tau
_{op}^{-1},$ which has only small kinks at frequencies related to the three
oscillators of the model $\alpha ^{2}F(\omega )$ given by
Eq.~(\ref{three-freq-model}). Of course, it is well known that optical data
is encoded with the information on the phonons even in the infinite band
case. However, a second derivative of the data is required \cite{marsiglio98},
and this is normally not as accurate.

In fig.~\ref{fig-omlam-self1} we plot $\omega\lambda _{op}(\omega )$
(solid curves) and $-2\Sigma _{1}(\omega )$ (dashed curves) for
the same values of $\Gamma$ as those for fig.~\ref{fig-tau-self2}.
In this case both quantities become smaller as impurity scattering
increases in the frequency range shown. Note that all curves cross zero,
which is a characteristic of finite bands. The crossing frequency depends
on $\Gamma$ for both $\omega\lambda _{op}(\omega )$ and
$-2\Sigma _{1}(\omega )$, with the former dependence stronger. Note
that in fig.~\ref{fig-omlam-self1} the phonon structures are again much more
prominent in the self energy as compared with its optical counterpart.
%%%%%%%%%%%%%%%%%%%%%%%%%%%%%%%%%%%%%%%
\begin{figure}
\twofigures[width=6.5cm]{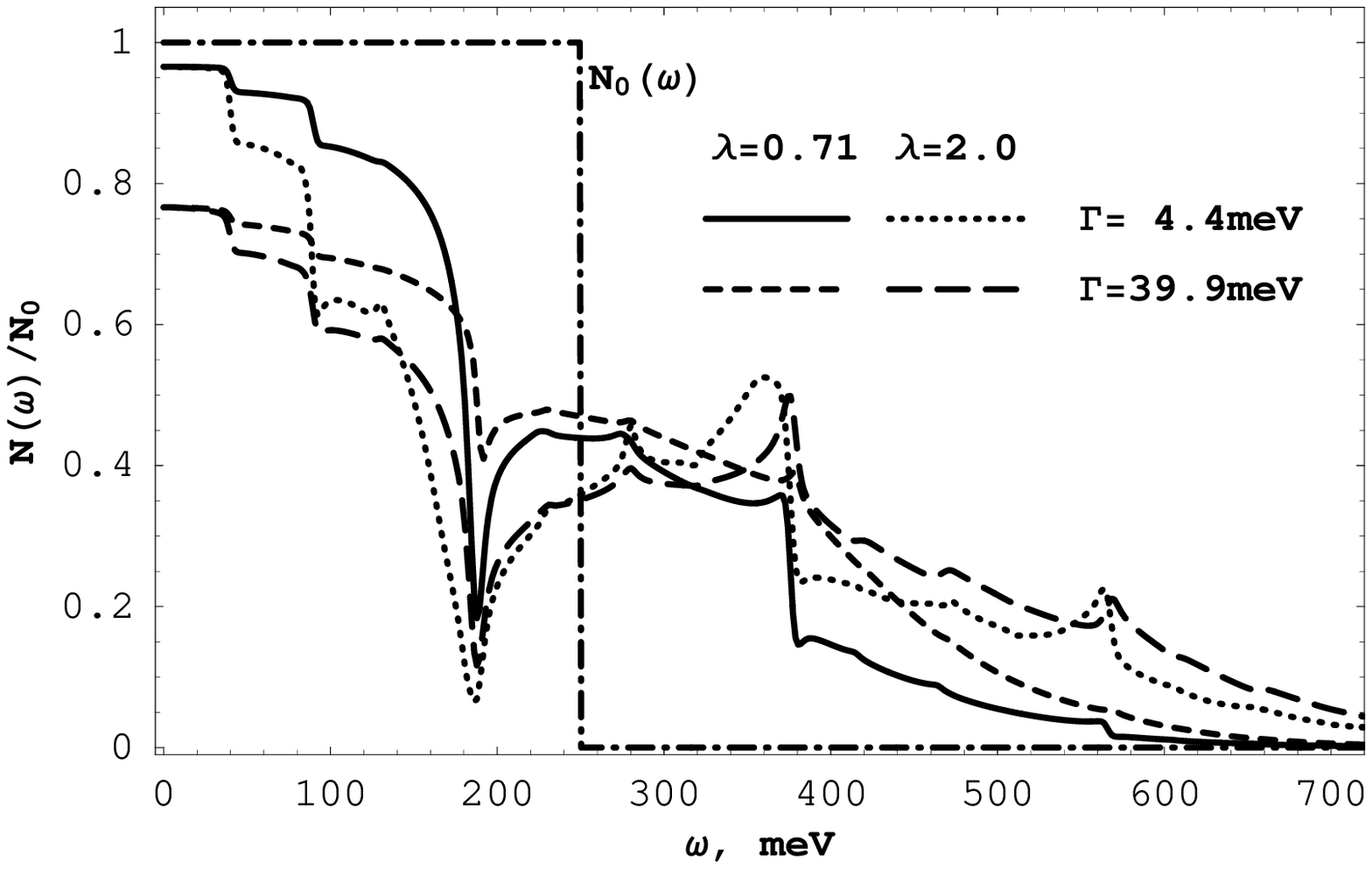}{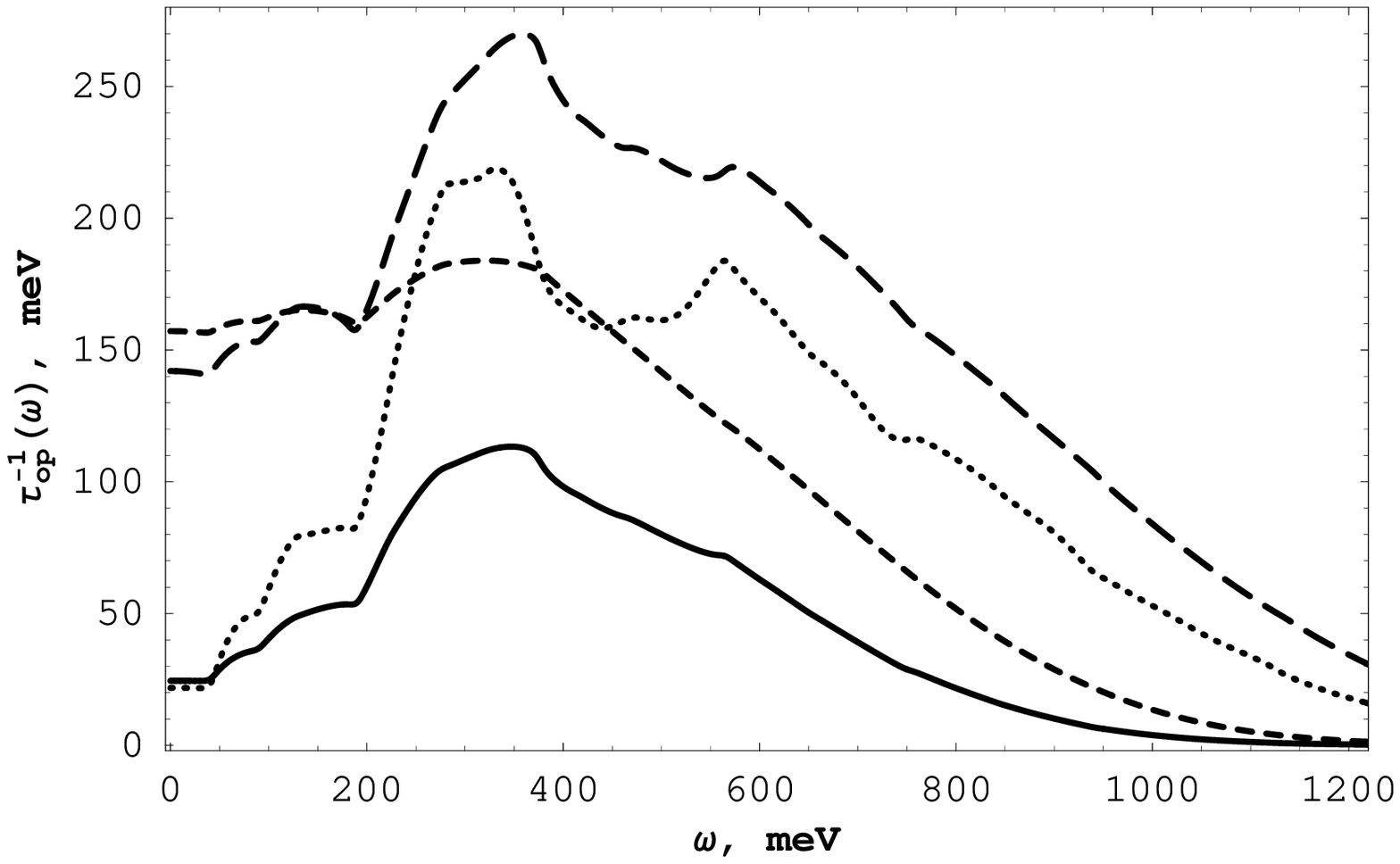}
%\twofigures[width=6.5cm]{fig-tau-self2-f.eps}{fig-omlam-self1-f.eps}
\caption{
The renormalized density of states $N(\omega )/N_0$ as a function
of frequency $\omega$ for the model $\alpha^2F(\omega)$ of
eq.~(\ref{three-freq-model}) relevant to $K_3C_{60}$. The curves
differ by
%impurity scattering rates and the
values of $\lambda$ and $\Gamma$ as indicated.
The dot--dashed curve is the bare EDOS with $W=0.5$ eV.
The temperature is $T=14.5$ K.
}
\label{fig-dos-fullerid}
\caption{
The optical scattering rate $\tau _{op}^{-1}(\omega )$ as a function
of frequency $\omega$ for the $K_3C_{60}$ model of
eq.~(\ref{three-freq-model}).
Parameters and marking of the curves are the same as in
fig.~\ref{fig-dos-fullerid}.
}
\label{fig-tau-fullerid}
\end{figure}
%%%%%%%%%%%%%%%%%%%%%%%%%%%
\section{Very narrow bands}
So far we have considered the case of a rather wide though finite electronic
band with full width $W = 2.5$ eV where $W\gg\omega_{\ln}$ and the phonon
structures in EDOS have a generic form.
However, band structure calculations for one family of the known
narrow materials --- fulleride compounds --- give a much smaller value
$W=0.5$ eV~\cite{satpathy92,gelfand94}. One expects that in this situation many
of the discussed features should be significantly enhanced \cite{dogan03}.
In fig.~\ref{fig-dos-fullerid} we present the renormalized EDOS $N(\omega )$
as a function of frequency $\omega $ at $T=14.5$ K for the model
$\alpha^2F(\omega)$ of eq.~(\ref{three-freq-model}) relevant to $K_3C_{60}$.
The bare EDOS was approximated by a flat EDOS with width $0.5$ eV,
and with $\mu =0$ (dot--dashed curve). The results shown are for impurity
scattering rates $\Gamma =4.4$ and $39.9$ meV, as indicated.
We used two values of the mass enhancement parameter:
$\lambda=0.71$~\cite{yoo00}, which is representative of the range quoted in the
literature~\cite{gunnarsson93,choi98}, and a bigger value $\lambda=2.0$, which
may be in a range more realistic for the fullerides~\cite{note1}.

In fig.~\ref{fig-dos-fullerid} we see that inclusion of the electron--phonon
interaction modifies the entire band, even for the modest value of $\lambda=0.71$.
The renormalized EDOS extends well beyond $\omega=W/2=250$ meV.
Note that what determines the energy of the highest steep drop in $N(\omega )$
is not the bare band edge but rather the energy of the third phonon mode
$\omega_3=190$ meV, which has the strongest coupling to band electrons in the
model of eq.~(\ref{three-freq-model}). Beyond this energy the multiphonon structures
appear, with magnitude depending both on $\lambda$ and on $\Gamma$; this extends the
renormalized band beyond $2.5\times(W/2)$ . On the other hand, no prominent feature
corresponding to the bare band edge remains in the renormalized EDOS.

The sharp drop in renormalized EDOS in fig.~\ref{fig-dos-fullerid}
at $\omega\approx\omega_3$  is in good quantitative agreement with photoemission
data on both polycrystalline \cite{gunnarsson93} and monolayer samples
\cite{yang03,yang05} of metallic fullerides. We note that the model of
$\alpha^2F(\omega)$ defined in eq.~(\ref{three-freq-model}) is expected to
work reasonably well even for two dimensional samples of fullerides because
the stronger coupled $\omega_3$ mode is associated mainly with intramolecular
vibrations of $C_{60}$ \cite{gunnarsson-rmp}. The flat bare EDOS  is also
appropriate for the two dimensional electonic bands. We conclude that
in a very narrow band metallic material the electron--phonon interactions
have a very profound impact on the density of electronic states observable by
various spectroscopies, to such an extent that they can dominate $N(\omega)$.

The importance of electron--phonon interactions and multiboson processes in
understanding photoemission data in fullerides has been emphasized by
Knupfer {\it et al.} \cite{gunnarsson93}. Using a different approach
they analysed the magnitude of $N(0)$ and redistibution of electronic states
derived from the bare band over frequency. In this letter we pointed out the
significance of finite band effects and provided a useful method to treat other
aspects of the problem, such as the effect of impurities, within the same formalism.
Sensitivity of these effects to the value of $\lambda$ could possibly be used
to extract the mass enhancement parameter for fulleride compounds from experimental
data on the EDOS, supplemented by the other spectroscopic techniques.
As an example, in fig.~\ref{fig-tau-fullerid} we show results for the optical
scattering rate in the $K_3C_{60}$ of eq.~(\ref{three-freq-model}) at $T=14.5$ K.
The curves shown are for $\lambda=0.71$ and $2.0$, and $\Gamma=4.4$ meV and
$39.9$ meV. They are marked the same way as in fig.~\ref{fig-dos-fullerid}.
Note that, similarly to the case of the EDOS, the phonon structures dominate the
frequency behavior of $\tau^{-1}_{op}(\omega)$ for very narrow band metals.

\acknowledgments
Work is supported by the Natural Science and Engineering Research Council of
Canada (NSERC) and the Canadian Institute for Advanced Research (CIAR).

%%%%%%%%%%%%%%%%%%%%%%%%%%%%%%%%%%%%%%%%%%%%%%%%%%%%%%%%5


\begin{thebibliography}{0}

\bibitem{mcmillan69} W.L. McMillan and J.M. Rowell, in {\it Superconductivity},
edited by R.D. Parks (Marcel Dekker, Inc., New York, 1969), p. 561.

\bibitem{steinrisser68} F. Steinrisser, L.C. Davis and C.B. Duke, Phys. Rev.
{\bf 176}, 912 (1968).

\bibitem{chen70} T.T. Chen and J.G. Adler, Solid State Commun. {\bf 8}, 1965 (1970).

\bibitem{wolf85} See, for example, E.L. Wolf,
{\it Principles of Electron Tunneling Spectroscopy},
(Oxford University Press, New York, 1985), for a thorough discussion.

%\bibitem{migdal58} A.B. Migdal, Zh. Eksp. Teor. Fiz. {\bf 34}, 1438 (1958)
%[Sov. Phys. JETP {\bf 7}, 996 (1958)].

\bibitem{englesberg63} S. Englesberg and J.R. Schrieffer, Phys. Rev.
{\bf 131}, 993 (1963).

\bibitem{mars-book}  F. Marsiglio and J.P. Carbotte,
in {\it The Physics of Superconductivity},
Vol. I: {\it Conventional and High T$_c$ Superconductors},
edited by K.H. Bennemann and J.B. Ketterson
(Springer Verlag, Berlin, 2003), p. 233.

\bibitem{mitrovic83} See, for example, a series of comprehensive
articles by B. Mitrovi\'{c} and J.P. Carbotte, Can. Jour. Phys.
{\bf 61}, 758 (1983); Can. Jour. Phys. {\bf 61}, 784 (1983);
Can. Jour. Phys. {\bf 61}, 872 (1983).

\bibitem{pickett82} W.E. Pickett, Phys. Rev. B {\bf 26}, 1186 (1982).

\bibitem{alexandrov87} A.S. Alexandrov, V.N. Grebenev, and E.A. Mazur,
Pis'ma Zh. Eksp. Teor. Fiz. {\bf 45} 357 (1987) [JETP Lett. {\bf 45}
455 (1987)].
%Note however that their interpetation of the features in EDOS is different
%from that suggested in Ref.~\cite{dogan03} and elaborated on here.
%%%
%Note however, that they actually took the presence of structure
%in the EDOS as a sign that the Migdal approximation broke down;
%that this conclusion was too strong was pointed out in Ref.~\cite{dogan03},
%where a more realistic phonon spectrum (not an Einstein spectrum) was used.

\bibitem{yoo00}  J.-W. Yoo and H.-Y. Choi, Phys. Rev. B {\bf 62}, 4440 (2000).

\bibitem{cappelluti01} E. Cappelluti, C. Grimaldi, and L. Pietronero,
Phys. Rev. B {\bf 64}, 125104 (2001).

\bibitem{cappelluti03}  E. Cappelluti and L. Pietronero, Phys. Rev. B {\bf 68},
224511 (2003). % [cond-mat/0309080].

\bibitem{dogan03}  F. Do\u{g}an and F. Marsiglio, Phys. Rev. B {\bf 68},
165102, (2003). % [cond-mat/0309075].

\bibitem{knigavko05}  A. Knigavko and J.P. Carbotte,
Phys. Rev. B {\bf 72}, 035125 (2005).

%\bibitem{dogan05} F. Do\u{g}an, unpublished.

\bibitem{marsiglio98}  F. Marsiglio, T. Startseva, and J.P. Carbotte,
Phys. Lett. A {\bf 245}, 172 (1998).

\bibitem{dordevic05}  S.V. Dordevic,
C.C. Homes, J.J. Tu, T. Valla, M. Strongin, P.D. Johnson, G.D. Gu, and D.N. Basov,
Phys. Rev. B{\bf 71}, 104529 (2005) %cond-mat/0411043.

\bibitem{marsiglio88}  F. Marsiglio, M. Schossmann, and J.P. Carbotte,
Phys. Rev. B {\bf 37}, 4965 (1988).

\bibitem{valla99}  T. Valla, A.V. Fedorov, P.D. Johnson, and S.L. Hulbert,
Phys. Rev. Lett. {\bf 83}, 2085 (1999).

%\bibitem{hengsberger99}  M. Hengsberger, R. Fresard, D. Purdie, P. Segovia,
%and Y. Baer, Phys. Rev. B {\bf 60}, 10796 (1999).

%\bibitem{lashell00} S. LaShell, E. Jensen, and T. Balasubramanian,
%Phys. Rev. B{\bf 61} 2371 (2000).

\bibitem{chainani00}  A. Chainani, T. Yokoya, T. Kiss, and S. Shin, Phys. Rev.
Lett. {\bf 85}, 1966 (2000).

\bibitem{reinert03}  F. Reinert,
B. Eltner, G. Nicolay, D. Ehm, S. Schmidt, and S. Hufner,
Phys. Rev. Lett. {\bf 91}, 186406 (2003).

%\bibitem{shi04} J. Shi {\it et al.},
%S.-J. Tang, B. Wu, P.T. Sprunger, W.L. Wang, V. Brouet,
%X.J. Zhou, Z. Hussain, Z.-X. Shen, Z. Zhang, and E.W. Plummer,
%Phys. Rev. Lett. {\bf 92}, 186401 (2004).

\bibitem{devereaux04} T.P. Devereaux, T. Cuk, Z.-X. Shen, and N. Nagaosa,
Phys. Rev. Lett. {\bf 93}, 117004 (2004).

\bibitem{choi98}  H.-Y. Choi, Phys. Rev. Lett. {\bf 81}, 441 (1998).

\bibitem{norman-nature}  M.R. Norman,
% {\it et al.},
H. Ding, M. Randeria, J.C. Campuzano, T. Yokoya, T. Takeuchi,
T. Takahashi, T. Mochiku, K. Kadowaki, P. Guptasarma, and D.G. Hinks,
Nature, {\bf 382}, 157 (1998).

\bibitem{marsiglio90}  F. Marsiglio and J.E. Hirsch,
Physica C {\bf 165}, 71 (1990).

\bibitem{satpathy92}  S. Satpathy,
% {\it et al.},
V.P. Antropov, O.K. Andersen, O. Jepsen, O. Gunnarsson, and A.I. Liechtenstein,
Phys. Rev. B {\bf 46}, 1773 (1992).

\bibitem{gelfand94}  M.P. Gelfand, Supercond. Review {\bf 1}, 103 (1994).

\bibitem{gunnarsson93} M. Knupfer,
% {\it et al.},
M. Merkel, M.S. Golden, J. Fink, O. Gunnarsson, and V.P. Antropov,
Phys. Rev. B {\bf 47}, R13944 (1993).

\bibitem{note1}
For $W\approx\omega_{\ln}$ the value of $\lambda$ extracted from
electronic self energy using the infinite band relation
$\lambda=-{\rm d}\Sigma_1(\omega=0)/{\rm d}\omega$ can be noticably smaller
than the one calculated from $\alpha^2F(\omega)$~\cite{cappelluti03}.


%\bibitem{benning93} P.J. Benning {\it et al.} %F. Stepniak, D.M. Poirier,
%J.L. Martins, J.H. Weaver, L.P.F. Chibante, and R.E. Smalley,
%Phys. Rev. B {\bf 47}, 13843 (1993).

\bibitem{yang03}  W.L. Yang,
%{\it et al.},
V. Brouet, X.J. Zhou, S.G. Louie, M.L. Cohen, S.A. Kellar, P.V. Bogdanov,
A. Lanzara, A. Goldoni, F. Parmigiani, Z. Hussain, and Z.-X. Shen,
Science {\bf 300}, 303 (2003).

\bibitem{yang05}  W.L. Yang, V. Brouet, X.J. Zhou, Z. Hussain,
and Z.-X. Shen, unpublished [2005 APS March meeting, report Y13.00006].

\bibitem{gunnarsson-rmp} O. Gunnarsson, Rev. Mod. Phys. {\bf 69}, 575 (1997).

\end{thebibliography}
\end{document}